\pdfoutput=1

\documentclass{appolb}

\usepackage{graphicx}

\usepackage{amsmath}
\usepackage{type1cm}
\usepackage{xspace}

\newcommand{\D}[0]{\mathrm{d}}
\newcommand{\Li}[0]{\mathrm{Li_2}}

\newcommand{\abs}[1]{\lvert #1 \rvert}
\newcommand{\Axiloop}[0]{\texttt{Axiloop}\xspace}
\newcommand{\del}[0]{\delta}
\newcommand{\eps}[0]{\epsilon}

\newcommand{\sdot}[2]{\,#1\!\cdot\!#2}

\begin{document}

\title{
\vspace{-4.0cm}
\begin{flushright}
		{\small {\bf IFJPAN-IV-2013-18}}   \\
		{\small {\bf SMU-HEP-13-24}}  \\
\end{flushright}
\vspace{0.5cm}
Virtual Corrections to the NLO Splitting Functions for Monte
  Carlo: the non-singlet case 
  \thanks{Presented by O.~Gituliar at Matter to The Deepest
  Conference, Ustro\'n, Poland,\\1--6~September~2013.
This work is partly supported by 
 the Polish National Science Center grant DEC-2011/03/B/ST2/02632,
  the Research Executive Agency (REA) of the European Union 
  Grant PITN-GA-2010-264564 (LHCPhenoNet),
the U.S.\ Department of Energy
under grant DE-FG02-13ER41996 and the Lightner-Sams Foundation.
          }
} 

\author{O.\ Gituliar$^a$, S.\ Jadach$^a$, A.~Kusina$^b$, M.\ Skrzypek$^a$
\address{$^a$ Institute of Nuclear Physics, Polish Academy of Sciences,\\
              ul.\ Radzikowskiego 152, 31-342 Cracow, Poland}
\address{$^b$ Southern Methodist University, Dallas, TX 75275, USA}
}

\maketitle

\begin{abstract}
Construction of a QCD cascade at the NLO level requires recalculation
of the splitting functions in a different manner \cite{Jadach:2013dta}.
We describe the calculation of
some of the virtual contributions to the non-singlet splitting function. In
order to be
compatible with the earlier calculated real contributions
\cite{Jadach:2011kc}, the 
principal value prescription for regularizing the infrared singularities must be
used in a new way. We illustrate this new scheme on simple examples.
For the calculations we wrote a Mathematica package called \Axiloop. We
describe its current status.
\end{abstract}



\section{Introduction}

With the second, 14 TeV, phase of the LHC experiments approaching, the need for
the precision QCD parton shower increases.
To date there are successful approaches of merging NLO hard process and
LO cascade like MC@NLO  \cite{Frixione:2002ik}
or POWHEG \cite{Nason:2004rx,Frixione:2007vw}.
Other attempts are also taken to improve precision of parton showers 
\cite{Tsuno:2006cu,Krauss:2005nu,Hoeche:2012yf,Alioli:2012fc, Hamilton:2012np}.
However, in order to construct QCD parton shower that
includes NNLO hard process, it will be mandatory to construct a
cascade at the NLO level.
Such a cascade is developed within the KRKMC
project~\cite{Skrzypek:2011zw,Jadach:2012vs, Jadach:2011cr}.

One of the crucial elements of this project is the recalculation and
reorganization of the NLO splitting functions.
The real emission part of the non-singlet (NS) splitting function has been 
discussed at length in \cite{Jadach:2011kc}.
The virtual $C_F^2$ NS components have been briefly discussed in
\cite{Gituliar:2013rta}.
Here we concentrate on the remaining NS virtual corrections.
We will also give an update on the development of the Mathematica package
\Axiloop \cite{axiloop},
that is written to assist with the NLO calculations in the
light-cone gauge.

\begin{figure}[h]
  \centerline{
  \includegraphics[height=2.5cm]{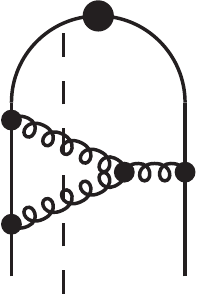}
  }
  \caption{The real graph $"(d)"$ contributing to the NLO non-singlet splitting
function.}
  \label{gr_d}
\end{figure}

Before going into details let us comment on the real corrections.
Let us consider a graph shown in Fig.\ \ref{gr_d}.
This interference graph contributes to the NS splitting function and is
labelled $"(d)"$.
By comparing the results for this graph from \cite{Jadach:2011kc} with the
previous ones of
\cite{Hei98} we note that they differ. 
The singular part of the graph "$(d)$" in \cite{Hei98}
reads (Table 3.10 p.\  99)
\begin{multline} \label{d_hei}
  \frac{p_{qq}(x)}{\eps^2}
  +
  2 I_0 \frac{p_{qq}(x)}{\eps}
  -
  \frac{1-x}{\epsilon}
  \\
  +
  p_{qq}(x)
  \Bigl(
    - 2 I_1 + 4 I_0 + 2 I_0 \ln{x} - 2 I_0 \ln(1-x)
  \Bigr)
  -
  2 (1-x) I_0
\end{multline}
and contains single and double poles in $\eps$ (the first line) and pure
spurious poles (the second line); 
where $p_{qq}(x) = (1+x^2)/(1-x)$ is the LO quark-quark splitting function,
$I_0\sim -\ln \delta$ and $I_1\sim -(1/2)\ln^2 \delta$ are the infrared
divergent functions regulated with $\delta$, cf.\ eq.\ (\ref{plus}), and
dimension of the phase space is defined as $m=4-2\epsilon$.
In \cite{Jadach:2011kc} for the same contribution we find only pure
spurious pole terms (eq.\ 3.48): 
\begin{equation} \label{d_stj}
  p_{qq}(x)
  \Bigl(
    2 I_1 + 4 I_0 + 2 I_0 \ln{x} - 2 I_0 \ln(1-x)
  \Bigr).
\end{equation}
Note, that those are semi-inclusive results which means that additional
integration over one real momentum is left to be calculated.
It is of the generic form 
$N(\epsilon)\int_0^{Q^2} d(-q^2) (-q^2)^{-1-2\epsilon}$. 
Such an integration introduces additional pole in $\eps$, so the inclusive
form of eq.\ (\ref{d_hei}) contains $1/\epsilon^3$ pole, whereas inclusive
form of eq.\ (\ref{d_stj}) contains just $1/\epsilon$ terms.%
\footnote{
Note, that in the eq.\ (3.48) of \cite{Jadach:2011kc} instead of $1/\epsilon$
pole one finds $\ln Q/q_0$ where $q_0$ is the lower limit of the integral 
$\int_{q_0}^{Q^2} d(-q^2) (-q^2)^{-1-2\epsilon}$, see eq.\ (2.10) of
\cite{Jadach:2011kc}.
         }.

Summarizing, the result (\ref{d_hei}) contains higher order poles in
$\epsilon$, absent in the new result in eq.\ (\ref{d_stj}). 
As explained in
\cite{Jadach:2013dta}, absence of higher order poles is
important for the construction of the stochastic cascade, which must be done in
four  dimensions.
Some clarifications are in order here:
(i) in the standard, inclusive, approach one adds real and virtual corrections
and, as can be seen in \cite{Hei98}, the higher order poles
cancel out. 
However, from the Monte Carlo point of view real-virtual cancellations pose
additional  complications.
(ii) The $1/\epsilon^2$ pole {\em does not} disappear in new
approach of \cite{Jadach:2011kc} --
the singularity is simply regulated differently, as $(1/\epsilon)\ln\delta$
etc., see below. 
(iii) Due to the presence of higher order poles, the ``trivial'' terms, $\ln
Q^2$, $\ln 4\pi$, etc., cancel between real and
virtual corrections in \cite{Hei98} and are
absent in \cite{Jadach:2011kc}. 
(iv) The complete NLO NS kernel, which is a coefficient of the single pole
is also affected by the
changes in higher order pole terms, however, it must be independent of the
choice of
the regularization. 
Therefore we expect that the difference between results (\ref{d_hei}) and
(\ref{d_stj}) will be compensated by an appropriate change in the virtual
contributions. 

Let us now discuss the mechanism responsible for this difference.


\section{Regularization Prescription}

Besides the fact that Lorentz structure of Feynman rules in the light-cone
gauge is more complicated than in the covariant gauges, additional difficulties
arise due to the specific structure of the axial denominator, 
$1/l_+$, where  $l_+= nl$ and $n$ is the axial vector. 
It leads to the spurious (non-physical) singularities when integrated over
virtual or real phase space. 
Therefore one must apply some prescription to regularize them in the
intermediate steps. 

The common choice is a principal value (PV) prescription \cite{Curci:1980uw,Ellis:1996nn,Hei98,Jadach:2011kc}.
It was originally used in \cite{Curci:1980uw} in the first calculation of the
NLO  splitting functions from the Feynman diagrams.
The other option is the Mandelstam-Leibbrandt prescription
\cite{Mandelstam:1982cb, Leibbrandt:1983pj}.
It is better founded in the QFT, but significantly more complicated in
practical  calculations.
Further analysis and comparison of these prescriptions can be found e.g.\ in
\cite{Hei98}.

The idea of the PV regularization is to replace $1/l_+$ terms by 
\begin{align} 
\frac{1}{l_+} \to \biggl[\frac{1}{l_+}\biggr]_{PV} 
         = \frac{l_+}{l_+^2 +\del^2p_+^2},
\label{plus}
\end{align}
where $\delta$ is an infinitesimal regulator, $p_+ =np$ and $p$ is some
reference momentum. 
This prescription has to be applied to the axial propagators of the gluons
\cite{Hei98} at the level of Feynman rules, i.e.\ at the very beginning of the
calculation, and it leads to the results like (\ref{d_hei}). 

Closer inspection of the derivations in \cite{Jadach:2011kc} reveals, that
there are, however, also different sources of the $1/l_+$ terms, related to the 
phase space evaluation, change of variables etc.
In the standard PV prescription these terms are {\em not} regularized by
$\delta$, dimensional regularization takes care of them.
On the contrary, we propose, that {\em all} these singularities in plus
variable {\em are} regulated also by the PV prescription.
In practice we replace 
\begin{align} 
l_+^{-1+\epsilon} \to \biggl[\frac{1}{l_+}\biggr]_{PV} l_+^{\epsilon}
\label{plusbar}
\end{align}
in {\em all} the places, keeping track of the higher order $\epsilon$ terms,
if needed.
Contrary to the previous case, this has to be done at the very end of the
integrations.
On the technical level it means that the usual (PV) approach:
(i) use some parametrization technique (like Feynman or Schwinger
parametrization) to combine denominators;
(ii) perform momentum integration;
(iii) perform integration over the Feynman/Schwinger parameters; is not
suitable.
Instead one needs to perform integration over $l_+$ at the very end, after the
integration over $l_-$, $l_\perp$, and $z_i$ variables:
\begin{multline} \label{eq:param}
  \int \frac{\D^m l}{(2\pi)^m}
    \frac{f(l_+)}{(l+k_1)^2 \ldots (l+k_n)^2}
  \\ =
  \int \D l_+ f(l_+)
  \int_0^1 \D z_1 \ldots \D z_{n-1}
  \int \frac{\D l_- \D^{m-2} l_\perp}{(2\pi)^m}
    \frac{1}{(l^2+\sdot{l}{A} + B^2)^n}.
\end{multline}
For explicit formulae arranged in this way, see for example
\cite{Ellis:1996nn}. 

As a consequence, now also the non-axial integrals, initially free of axial
vector $n$ (which defines the plus component, $l_+=nl$), can develop dependence
on the vector $n$. 
For example, let us show the three-point integral without axial denominator
\begin{equation} \label{eq:def-feyn}
  J_3^\mathrm{F}
  =
  \int \frac{\D^m l}{(2\pi)^m}
    \frac{1}{l^2 (l+q)^2 (l+p)^2}
\end{equation}
for the special kinematic configuration: $p^2=(p-q)^2=0$.
Standard approach leads to the following expression:
\begin{equation}
  J_3^\mathrm{F}
  =
  \frac{i}{(4\pi)^2 \abs{q^2}}
  \left(
    \frac{4\pi}{\abs{q^2}}
  \right)^\eps
  \Gamma(1+\eps)
  \left(
    -\frac{1}{\eps^2} + \frac{\pi^2}{6}
  \right)
\end{equation}
In contrast our prescription gives the result:
\begin{multline} \label{eq:our-f}
  J_3^\mathrm{F}
  =
  \frac{i}{(4\pi)^2 \abs{q^2}}
  \left(
    \frac{4\pi}{\abs{q^2}}
  \right)^\eps
  \Gamma(1+\eps)
  \bigg(
    \frac{2 I_0 + \ln(1-x)}{\eps}
    \\
    - 4 I_1 + 2 I_0 \ln(1-x) + \frac{\ln^2(1-x)}{2}
  \bigg)
\end{multline}
where $x=q_+/p_+$ is the axial-vector-dependent parameter. 
As one can see, eq.\ \eqref{eq:our-f} 
is free of double poles in $\eps$.
Note that this singularity did not disappear, but has been replaced by
$I_1$ and $I_0/\epsilon$. 
More details will be presented elsewhere \cite{stjinprep}.

\section{Results}

\begin{figure}[h]
  \centerline{
    \includegraphics[height=2.25cm]{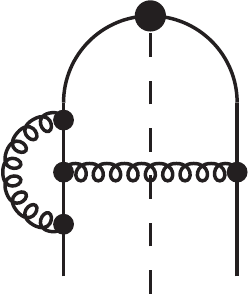}
    \hspace{0.15cm}
    \includegraphics[height=2.25cm]{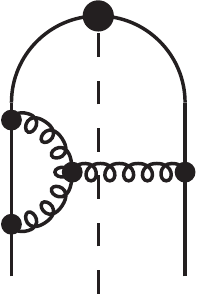}
    \hspace{0.3cm}
    \includegraphics[height=2.25cm]{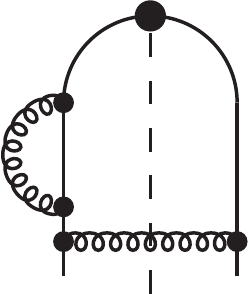}
    \hspace{0.3cm}
    \includegraphics[height=2.25cm]{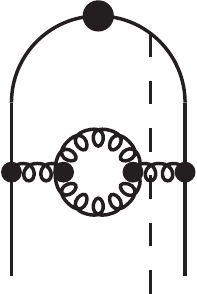}
    \hspace{0.3cm}
    \includegraphics[height=2.25cm]{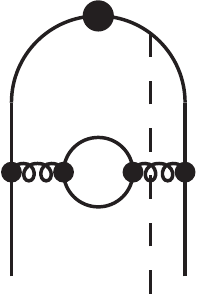}
  }
\hskip 2cm (c)\hskip 1.4cm (d)\hskip 2cm (e)\hskip 1.6cm (f)\hskip 1.6cm (g)
  \caption{A complete set of the non-singlet virtual contributions to
the NLO splitting function.}
  \label{fig:ns-set}
\end{figure}

Let us now use the new prescription in the actual calculation. We present
here results for one of the virtual graphs contributing to the non-singlet
splitting function, namely graph "$(d)$", i.e.\ the second graph in Fig.\
\ref{fig:ns-set}.
The presented results have been obtained with the help of the Mathematica
package \Axiloop, which we developed to assist with the calculations in
the light-cone gauge. Overview of the \Axiloop package is provided in the
Appendix.

In Table \ref{tb:ff} we show inclusive contributions separately for
the real graph "(d)" (Fig.\ \ref{gr_d}) taken from eq.\ (3.48) of
\cite{Jadach:2011kc} and virtual graph "(d)" (Fig.\ \ref{fig:ns-set}, second
picture),
calculated in the new regularization scheme.
\begin{table}
  \centering
  \begin{tabular}{|l|r|r|r|}
    \hline
    & \includegraphics[height=1.5cm]{nlo_d.pdf}
    & \includegraphics[height=1.5cm]{nlo_d_n_r.pdf}
    & SUM
    \\ \hline
    \multicolumn{4}{|c|}{Double poles}
    \\ \hline
        $p_{qq}$
      & $-{3}/{2}$
      & $0$
      & $-{3}/{2}$
    \\
        $p_{qq} \; I_0$
      & $4$
      & $0$
      & $4$
    \\ 
        $p_{qq} \; \ln{x}$
      & $1$
      & $0$
      & $1$
    \\ 
        $p_{qq} \; \ln(1-x)$
      & $2$
      & $0$
      & $2$
    \\ \hline
    \multicolumn{4}{|c|}{Single poles}
    \\ \hline
        $p_{qq}$
      & $-7$
      & $-4$
      & $-11$
    \\
        $1-x$
      & $-5/2$
      & ${3}/{2}$
      & $-1$
    \\
        $1+x$
      & $-{1}/{2}$
      & ${1}/{2}$
      & $0$
    \\
        $p_{qq} \; \ln{x}$
      & $ 0$
      & $-{3}/{2}$
      & $-{3}/{2}$
    \\
        $(1-x) \; \ln{x}$
      & $2$
      & $0$
      & $2$
    \\
        $(1+x) \; \ln{x}$
      & $0$
      & ${1}/{2}$
      & ${1}/{2}$
    \\
        $p_{qq} \; \ln(1-x)$
      & $-3$
      & $8$
      & $5$
    \\
        $(1-x) \; \ln(1-x)$
      & $4$
      & $0$
      & $4$
    \\
        $p_{qq} \; \ln^2{x}$
      & $2$
      & $-1$
      & $1$
    \\
        $p_{qq} \; \ln{x} \ln(1-x)$
      & $2$
      & $4$
      & $6$
    \\
        $p_{qq} \; \ln^2(1-x)$
      & $4$
      & $-2$
      & $2$
    \\
        $p_{qq} \; \Li(1-x)$
      & $-2$
      & $2$
      & $0$
    \\
        $p_{qq} \; \Li(1)$
      & $8$
      & $-2$
      & $6$
    \\ \hline
        $p_{qq} \; I_1$
      & $-12$
      & $4$
      & $-8$
    \\
        $p_{qq} \; I_0$
      & $0$
      & $8$
      & $8$
    \\
        $(1-x) \; I_0$
      & $8$
      & $0$
      & $8$
    \\
        $p_{qq} \; I_0 \ln{x}$
      & $4$
      & $4$
      & $8$
    \\
        $p_{qq} \; I_0 \ln(1-x)$
      & $12$
      & $-4$
      & $8$
    \\ \hline
  \end{tabular}
  \caption{
Contributions from real and virtual graphs "(d)" to inclusive splitting
function 
and their sum, calculated in the new regularization
prescription.}
 \label{tb:ff}
\end{table}
Table is constructed in analogy to Table 1 of \cite{Curci:1980uw} but in
addition 
contains a section with double pole contributions.

Let us describe these results in more detail.

First of all, the last column in Table \ref{tb:ff} is in full agreement
with the
result of Table 1 of \cite{Curci:1980uw}, where only
the sum of real and virtual contributions is given. The 
virtual contribution alone can be obtained from \cite{Hei98} in the standard
PV prescription. 

The second point is that $1/\eps^3$ pole is absent in the inclusive
virtual contribution, in a similar manner as it disappeared from the
inclusive real
contribution, see discussion of eq.~\eqref{d_stj}.
Results obtained by applying the PV prescription (see
eq.~\eqref{d_hei} for the real graph) do not have this property.

Thirdly, neither real nor virtual contribution depends on the
scale $Q$ of the hard process. This is not true for the standard
prescription -- in that case only the sum of real and virtual terms is
independent of $Q$. 
The same happens also for the "trivial" terms, like $\ln 4\pi$ or $\gamma_E$.
Note however, that 
there are some contributions for which this property does not hold.

\section{Summary and outlook}

In this note we presented some details of the recalculation of virtual graphs
contributing to the NLO NS splitting function, necessary for the
construction of the Monte Carlo cascade at the NLO level. We argued, that in
order to be compatible with the earlier calculation \cite{Jadach:2011kc} for
the real
components, the PV prescription must be used in a modified way.
As a result, at the inclusive level, the $1/\epsilon^3$ poles are replaced
by the structures like $\ln\delta/\epsilon^2$ and no cancellations of
$1/\epsilon^3$ poles between real and virtual parts is needed.
We performed calculations for the case of the NS splitting function
and presented them here on the example of one diagram. As expected, our
results for real and virtual contributions differ from the ones in PV
prescription, but the sum of real and virtual contributions is the same.%
\footnote{Let us note that the importance of the PV regularization 
on the way from the $\overline{MS}$ to the ``physical'' collinear factorization
scheme
is also underlined in the recent paper~\cite{deOliveira:2013maa}.
         }.
In the next step we plan to apply the new prescription to the singlet virtual
graphs. 

In order to automatize the calculations in the light-cone gauge we developed
a Mathematica package \Axiloop with the help of which a set of one-loop graphs
contributing to the NS
splitting functions at next-to-leading order can be calculated
in the standard and modified PV prescriptions.
In the future the package can be extended to calculate singlet
one-real-one-virtual as well as two-loop and two-real corrections to the
splitting functions.

\section*{Acknowledgment}
This work is partly supported by 
 the Polish National Science Center grant DEC-2011/03/B/ST2/02632,
  the Research Executive Agency (REA) of the European Union 
  Grant PITN-GA-2010-264564 (LHCPhenoNet),
the U.S.\ Department of Energy
under grant DE-FG02-13ER41996 and the Lightner-Sams Foundation.
Two of the authors (S.J. and M.S) are grateful for the
the warm hospitality of the TH Unit of the CERN PH Division,
while completing this work.

\section*{Appendix: Overview of the \Axiloop software package}
\label{appen}

For calculating virtual corrections to the NLO splitting
functions we built a software package for the Wolfram Mathematica
system --- \Axiloop \cite{axiloop}. 
In this section, we describe a general structure as well as some core
features of the package implemented up to date. 

\Axiloop is an open-source, general-purpose package which provides a
complete set of routines for calculating Feynman-diagram-based objects in
light-cone gauge in
analytical form. 
In the current version a full set of the virtual diagrams
contributing to the non-singlet NLO
splitting functions (Fig.~\ref{fig:ns-set}) is calculated using a new
regularization prescription. 

Functions provided by \Axiloop may be divided into the two main groups:
a set of general-purpose {\it core routines}, which likely can be
  used in other packages for solving general-purpose problems and 
{\it custom routines} dedicated to calculation of the
splitting functions.

{\it Core routines} perform the following tasks:
trace and vector algebra operations in arbitrary number of
dimensions~\cite{Jamin:1991dp};
virtual integrals in light-cone and Feynman gauges with custom
regularization schemes (including the one discussed in this paper);
various simplification algorithms for loop integrals 
(e.g.\ \cite{Capper:1981rc}); 
reduction of tensor integrals using  Passarino-Veltman approach
\cite{Denner:2005nn}; 
final-state integration.
It is worth to emphasize that 
\Axiloop provides a flexible implementation of the loop integration routines
which can be modified and extended for various contexts.

One of the core routines is \texttt{IntegrateLoop} function which performs
one-loop integration in $m=4-2\eps$ dimensions.
It handles Feynman and axial two- and three-point integrals with up to a rank 3
tensor structures in the numerator.
As a demonstration we show the scalar
integral of eq.~\eqref{eq:def-feyn}, which in \Axiloop notation looks as
follows:
\begin{verbatim}
In[1]:= IntegrateLoop[ 1/(l.l (l+k).(l+k) (l+p).(l+p)), l]
Out[1]= Q[eps] (-k.k)^(-1-eps) ((2 I0 + Log[1-x])/eir
        - 4 I1 + 2 I0 Log[1-x] + Log[1-x]^2/2)
\end{verbatim}
In general, we distinguish infra-red and ultra-violet singularities when
dimensional regularization prescription is used.
The infrared origin of the pole is indicated by the \texttt{"eir"} symbol.
Ultra-violet poles, which appear in two- and some three-point integrals, are
represented by poles in the \texttt{"euv"} symbol.

{\it Custom routines} in our case are dedicated to the calculation of the NLO
virtual splitting functions. 
They produce analytical expressions for the components of the splitting
functions at different stages of the calculation, as defined in
\cite{Gituliar:2013rta}:
  (i)~renormalized and bare exclusive formulas, which are usually
omitted by other authors -- they play a key role in the construction of the
parton shower Monte Carlo; 
  (ii)~inclusive results for cross-checking with previous known results; and
  (iii)~ultra-violet counter-terms.

The main routine for calculating splitting functions is
\texttt{SplittingFunction}. 
As its input a complete description of the calculated graph in
terms of the Feynman rules should be provided. 
The following example demonstrates invocation of \texttt{SplittingFunction} for
calculating virtual contribution of the topology "(d)": 
\begin{verbatim}
In[3]:= SplittingFunction[ G[n]/(4 p.n) FP[k] FV[i1] FP[l] FV[i2]
  GP[i1,i3, l+k] GP[i2,i4, l+p] GV[i3,-l-k, i4,l+p, mu,-p+k]
  FPx[p] GPx[mu,nu, p-k] FV[nu] FP[k]
  ,
  IntegrateLoopPrescription -> "MPV"
]
\end{verbatim}
The \texttt{IntegrateLoopPrescription} option allows one to change
regularization prescription for loop integrals. 
It accepts \texttt{"PV"} and \texttt{"MPV"} values for the PV prescription 
or its modification (described in this work) respectively.
Remaining routines, i.e. \texttt{FP} (\texttt{FPx}), \texttt{GP}
(\texttt{GPx}), \texttt{FV}, and \texttt{GV}, correspond to the Feynman rules
and read as fermion/gluon propagator (suffix \texttt{"x"} indicates that
corresponding propagator is cut) and fermion/gluon vertex respectively. 

At the moment in the \Axiloop package we have implemented all the NS
one-real-one-virtual corrections
 (corresponding graphs
are depicted in the
Fig.~\ref{fig:ns-set}) and the library of two- and three-point integrals in
the light-cone gauge for both regularization schemes.


\providecommand{\href}[2]{#2}
\end{document}